\documentclass[amsmath,amssymb,12pt,aps]{revtex4-2}
\usepackage{amsfonts}
\usepackage[utf8]{inputenc} 
\usepackage{graphicx}
\usepackage{float}
\usepackage{natbib}
\usepackage{wrapfig}
\usepackage{xcolor}
\usepackage{dcolumn}
\usepackage[english]{babel}
\begin{document}
	
\title{Exploring the phase diagrams of multidimensional Kuramoto models}

\author{Ricardo Fariello}

\affiliation{Departamento de Ci\^encias da Computa\c{c}\~ao,
	Universidade Estadual de Montes Claros, 39401-089, Montes Claros, MG, Brazil.} 
	
\author{Marcus A. M. de Aguiar }

\affiliation{Instituto de F\'isica Gleb Wataghin, Universidade Estadual de Campinas, Unicamp 13083-970, Campinas, SP, Brazil}

\begin{abstract}
	
The multidimensional Kuramoto model describes the synchronization dynamics of particles moving on the surface of D-dimensional spheres, generalizing the original model where particles were characterized by a single phase. In this setup, particles are more easily represented by $D$-dimensional  unit vectors than by $D-1$ spherical angles, allowing for the coupling constant to be extended to a coupling matrix acting on the vectors. As in the original Kuramoto model, each particle has a set of $D(D-1)/2$ natural frequencies, drawn from a distribution. The system has a large number of independent parameters, given by the average natural frequencies, the characteristic widths of their distributions  plus $D^2$ constants of the coupling matrix. General phase diagrams, indicating regions in parameter space where the system exhibits different behaviors, are hard to derive analytically. Here we obtain the complete phase diagram for $D=2$ and Lorentzian distributions of natural frequencies using the Ott-Antonsen ansatz. We also explore the diagrams numerically for different distributions and some specific choices of parameters for $D=2$, $D=3$ and $D=4$. In all cases the system exhibits at most four different phases: disordered, static synchrony, rotation and active synchrony.  Existence of specific phases and boundaries between them depend strongly on the dimension $D$, the coupling matrix and the distribution of natural frequencies.
	
\end{abstract}

\maketitle

% %%%%%%%%%%%%%%%%%%%%%%%%%%%%%%%%%%%%%%%%%%%%%%%%%%%%
% %%%%%%%%%%%%%%%%%%%%%%%%%%%%%%%%%%%%%%%%%%%%%%%%%%%%
\section{Introduction}

Many natural and artificial systems can be described mathematically by a set of coupled oscillators. Examples include neuronal networks \cite{cumin2007generalising,bhowmik2012well,ferrari2015phase,reis2021bursting}, power grids \cite{filatrella2008analysis,motter2013spontaneous,Nishikawa_2015,molnar2021asymmetry}, active matter \cite{han2020emergence}, sperm motion \cite{riedel2005self,wright2023thermodynamics}, coupled metronomes \cite{Pantaleone2002} and  circadian rhythms \cite{yamaguchi2003,bick2020understanding}. In all these examples, synchronous motion of the oscillators is a key feature, leading  to macroscopic behaviors with important consequences. The model introduced by Kuramoto provided the first detailed study of synchronization in a simple setup, becoming a paradigm in the area \cite{Kuramoto1975,Kuramoto1984}. In this model the oscillators are represented only by their phases $\theta_i$ and evolve according to the equations 
\begin{equation}
	\dot{\theta}_i = \omega_i + \frac{k}{N} \sum_{j=1}^N \sin{(\theta_j-\theta_i)}
	\label{kuramoto}
\end{equation}
where  $\omega_i$ are their natural frequencies, selected from a symmetric distribution $g(\omega)$, $k$ is the coupling strength and $i = 1, ..., N$.  Kuramoto showed that, for $k$ is sufficiently large, the oscillators synchronize their phases. A measure of synchronization is given by the complex order parameter 
\begin{equation}
	z = p e^{i \psi} \equiv \frac{1}{N} \sum_{i=1}^N e^{i\theta_i}
	\label{paraord}
\end{equation}
which is $p \approx 0$ for independent oscillators and $p \approx 1$ when motion is coherent. In the limit $N \rightarrow \infty$, the onset of synchronization can be described as a continuous phase transition, where $p=0$ for $k < k_c =  2/\pi g(0)$ and increases as $p = \sqrt{1-k_c/k}$ for $ k > k_c$ \cite{Acebron2005,Rodrigues2016}. 

Since its original inception, the model was extended in many ways, with the introduction of frustration \cite{sakaguchi1986soluble,yue2020model,buzanello2022matrix,de2023generalized}, 
different types of coupling functions \cite{hong2011kuramoto,yeung1999time,breakspear2010generative}, networks \cite{Rodrigues2016,Joyce2019}, distributions of the oscillator's natural frequencies  \cite{Gomez-Gardenes2011,Ji2013}, inertial terms \cite{Acebron2005,dorfler2011critical,olmi2014hysteretic}, external periodic driving forces \cite{Childs2008,moreira2019global,moreira2019modular} and coupling with particle swarms \cite{o2017oscillators,o2022collective,supekar2023learning}.  

The Kuramoto model was also extended to higher dimensions with the help of the unit vectors $\vec{\sigma_i} = (\cos{\theta_i},\sin{\theta_i}) \equiv (\sigma_{ix},\sigma_{iy})$ 
\cite{chandra2019continuous}.  Computing $\dot{\sigma}_{ix} = - \dot{\theta}_i \sigma_{iy}$,  $\dot{\sigma}_{iy} = \dot{\theta}_i \sigma_{ix}$ and using Eq.(\ref{kuramoto}) it follows that
\begin{equation}
	\frac{d \vec{\sigma_i}}{d t} = \mathbf{W}_i \vec{\sigma_i} + \frac{k}{N} \sum_j [\vec{\sigma_j} - (\vec{\sigma_i}\cdot \vec{\sigma_j}) \vec{\sigma_i}]
	\label{eq3}
\end{equation}
where  $\mathbf{W}_i$ is the anti-symmetric matrix 
\begin{equation}
	\mathbf{W}_i = \left( 
	\begin{array}{cc}
		0 & -\omega_i \\
		\omega_i & 0
	\end{array}
	\right).
	\label{wmat}
\end{equation}
The complex order parameter  $z$, Eq.(\ref{paraord}), can be written in terms of the real vector
\begin{equation}
	\vec{p} = \frac{1}{N}\sum_i \vec{\sigma_i} = (p\cos\psi,p\sin\psi)
	\label{vecpar}
\end{equation}
describing the center of mass of the system. 

Eq.(\ref{eq3}) can be extended to higher dimensions by simply considering unit vectors $\vec\sigma_i$ in D-dimensions, rotating on the surface of the corresponding (D-1) unit sphere 
\cite{chandra2019continuous}. Particles are now represented by $D-1$ spherical angles, generalizing the single phase $\theta_i$ of the original model. The matrices $\mathbf{W}_i$ become $D \times D$ anti-symmetric matrices containing the $D(D-1)/2$ natural frequencies of each oscillator. Finally, the $D$-dimensional model is further extended by replacing the coupling constant $k$ by a coupling matrix $\mathbf{K}$ acting on the vectors \cite{barioni2021complexity,buzanello2022matrix,de2023generalized}:
\begin{equation}
	\frac{d \vec{\sigma_i}}{d t} = \mathbf{W}_i \vec{\sigma_i} + \frac{1}{N} \sum_j [{\mathbf K} \vec{\sigma_j} - (\vec{\sigma_i}\cdot {\mathbf K} \vec{\sigma_j}) \vec{\sigma_i}] .
	\label{kuragen}
\end{equation}
Using Eq.(\ref{vecpar}) and defining the {\it rotated} order parameter
\begin{equation}
	\vec{q} = \mathbf{K} \vec{p}
	\label{qpar}
\end{equation}
we obtain the compact equation
\begin{equation}
		\frac{d \vec{\sigma_i}}{d t} =  \mathbf{W}_i \vec{\sigma_i} +  [ \vec{q}- (\vec{\sigma_i}\cdot \vec{q}) \vec{\sigma_i}].
	\label{kuragen3}
\end{equation}
The coupling matrix breaks the rotational symmetry and plays the role of a generalized frustration: it rotates $\vec\sigma_j$, hindering its alignment with $\vec\sigma_i$ and inhibiting synchronization.  The angle of rotation depends on $\sigma_j$, generalizing the constant frustration angle of the Sakaguchi model \cite{sakaguchi1986soluble}.  Norm conservation, $|\vec{\sigma_i}|=1$, is guaranteed, as can be seen by taking the scalar product of Eqs.(\ref{kuragen}) with $\vec{\sigma_i}$.  Similar extensions of the Kuramoto model with symmetry breaking and higher dimensions were also considered in  refs. \cite{Tanaka2014,lipton2021kuramoto,crnkic2021synchronization,manoranjani2023diverse,lee2023chimera}.

For the case of identical oscillators with zero natural frequencies, the eigenvalues and eigenvectors of the coupling matrix completely 
determine the dynamics \cite{de2023generalized}. Complete synchronization occurs if the real part of the dominant eigenvalue is positive.
If the corresponding eigenvector is real the order parameter converges to the direction of the eigenvector (static sync). If it is complex,
the order parameter rotates in the plane defined by the real and imaginary parts of the corresponding eigenvector. However, for non-identical
oscillators, the behavior changes considerably, depending on the distribution of natural frequencies. Not only the center $\omega_0$ 
and width $\Delta$ of the distribution matter (since rotational symmetry is broken) but also the type of distribution. Here we construct phase 
diagrams in the $\omega_0 \times \Delta$ plane for different distributions and dimensions 2, 3 and 4,  exploring the effects of these
parameters in the dynamics of the extended Kuramoto model.

% %%%%%%%%%%%%%%%%%%%%%%%%%%%%%%%%%%%%%%%%%%%%%%%%%%%%
% %%%%%%%%%%%%%%%%%%%%%%%%%%%%%%%%%%%%%%%%%%%%%%%%%%%%
\section{Representations in 2, 3 and 4 dimensions}
\label{rep}

In $D=2$ the coupling matrix $\mathbf{K}$ can be conveniently written as a sum of symmetric and anti-symmetric parts
\begin{equation}
	\mathbf{K} = K
	\left( 
	\begin{array}{cc}
		\cos\alpha & \sin\alpha \\
		-\sin\alpha & \cos\alpha
	\end{array}
	\right) + 
	J \left( 
	\begin{array}{cc}
		-\cos\beta & \sin\beta \\
		\sin\beta & \cos\beta
	\end{array}
	\right) \equiv \mathbf{K}_R  + \mathbf{K}_S  
	\label{k2D}
\end{equation}
where $\mathbf{K}_R$ is recognized as a rotation matrix. In this case the equations of motion can still be written in terms of
a single phase and read
\begin{equation}
	\dot{\theta}_i = \omega_i + \frac{1}{N} \sum_{j=1}^N \left[ K \sin(\theta_j - \theta_i - \alpha) + J \sin(\theta_j + \theta_i + \beta) \right].
	\label{kuragenjk}
\end{equation}
For $J=0$ the system reduces to the Kuramoto-Sakaguchi model, but for $J \neq 0$ new {\it active} states are obtained \cite{buzanello2022matrix}. We
review the main properties of the 2D system in the next section.

As the coupling matrix has $D^2$ independent real entries, the model equations are hard to handle explicitly if $D \geq 3$. For identical oscillators, 
$\mathbf{W}_i=0$, the dynamics  is completely determined by the eigenvalues and eigenvectors of $\mathbf{K}$ \cite{de2023generalized}, but for 
general distributions of natural frequencies the dynamics changes considerably, and so does the phase diagram of the system in the space of parameters.
In order to simplify matters, we choose to work with particular forms of the coupling matrices that make it easy to identify the leading eigenvectors and,
therefore, to predict the behavior of the system in the limit where the width of the distribution of natural frequencies goes to zero.

For $D=3$ we set
\begin{equation}
	\mathbf{K} = \left(
	\begin{array}{ccc}
		a \cos \alpha & a \sin \alpha & 0 \\
		-a \sin \alpha & a \cos \alpha & 0 \\
		0 & 0 & b
	\end{array}
	\right).
	\label{3d}
\end{equation}
The eigenvalues are $\lambda_\pm = a e^{\pm i\alpha}$, with eigenvectors $(1,\pm i,0)/\sqrt{2}$, and $\lambda_3=b$, with eigenvector $(0,0,1)$. 
The matrices of natural frequencies have three components each and are given by
\begin{equation}
	\mathbf{W}_i = \left( 
	\begin{array}{ccc}
		0 & -\omega_{3i}  &  \omega_{2i} \\
		\omega_{3i} & 0  &  -\omega_{1i} \\
		-\omega_{2i} & \omega_{1i} & 0 
	\end{array}
	\right).
	\label{wmat3}
\end{equation}
In this case it is possible to associate $\mathbf{W}_i$ to the vector
\begin{equation}
	\vec{\omega}_i^T = \omega_i (\sin\beta_i \cos \alpha_i, \sin\beta_i \sin\alpha_i,\cos\beta_i),
	\label{vecwmat3}
\end{equation}
where the superscript $T$ stands for transpose. Clearly $\mathbf{W}_i \vec{\sigma} = \omega_i \times \vec{\sigma}$.

For $D=4$ we choose the coupling matrix as
\begin{equation}
	\mathbf{K} = \left(
	\begin{array}{cccc}
		a_1 \cos \alpha & a_1 \sin \alpha & 0 & 0\\
		-a_2 \sin \alpha & a_2 \cos \alpha & 0 & 0\\
		0 & 0 & b \cos \beta & b \sin \beta\\
		0 & 0 & -b \sin\beta & b \cos \beta 
	\end{array}
	\right),
	\label{4d}
\end{equation}
representing a rotation in the lower block and another rotation (if $a_1=a_2$) or two real eigenvectors (if $\alpha=0$ and $a_1 \neq a_2$) in the upper block. 
We note that any real coupling matrix could be used and that only the eigenvalues and eigenvectors matter for the asymptotic behavior of the system in
the case of identical oscillators. This choice is only to facilitate the determination of the eigenvectors. The matrices $\mathbf{W}_i$ can be parametrized as \cite{barioni2021ott}
\begin{equation}
	\mathbf{W}_i = \left( 
	\begin{array}{cccc}
		0 & -\omega_{6i}  &  \omega_{5i} & -\omega_{4i} \\
		\omega_{6i} & 0  &  -\omega_{3i}  & \omega_{2i}\\
		-\omega_{5i} & \omega_{3i} & 0 & -\omega_{1i}  \\
		\omega_{4i} & -\omega_{2i}  & \omega_{1i} & 0\
	\end{array}
	\right)
	\label{wmat4}
\end{equation}
and have six independent entries.

% %%%%%%%%%%%%%%%%%%%%%%%%%%%%%%%%%%%%%%%%%%%%%%%%%%%%
% %%%%%%%%%%%%%%%%%%%%%%%%%%%%%%%%%%%%%%%%%%%%%%%%%%%%
\section{Exact results for $D=2$}
\label{exact}

\subsection{Dimensional reduction}

In two dimensions we can use the dimensional reduction approach of Ott and Antonsen for Lorentzian
distributions of natural frequencies \cite{Ott2008}. In this case one can derive differential equations for the module and phase of the order
parameter \cite{buzanello2022matrix} and we will use these equations to construct the full 2D phase diagram analytically. Taking the limit $N \rightarrow \infty$ we define the function
$f(\omega,\theta,t)$ describing the density of oscillators with natural frequency $\omega$ at position $\theta$ in
time $t$. Since the total number of oscillators is conserved, $f$ satisfies a continuity equation with velocity field
\begin{equation}
	\vec{v} =  \mathbf{W} \vec{\sigma} +   \vec{q}- (\vec{\sigma}\cdot \vec{q}) \vec{\sigma} = [\omega + q \sin(\xi-\theta)] \hat{\theta} \equiv v_\theta \hat{\theta}
	\label{vtheta}
\end{equation}
where $\vec{\sigma}=(\cos\theta,\sin\theta)$, $\vec{q} \equiv q (\cos\xi,\sin\xi)$ and $\hat{\theta}=(-\sin\theta, \cos\theta)$.
The continuity equation reads
\begin{equation}
	\frac{\partial f}{\partial t} + \frac{\partial (v_\theta f)}{\partial \theta} = 0.
	\label{cont1}
\end{equation}

The Ott-Antonsen ansatz consists in expanding $f$ in Fourier series and choose the coefficients so that all of them depend on a single
complex parameter $\nu(t)$ as
\begin{equation}
	f(\omega,\theta,t) = \frac{g(\omega)}{2\pi} \left[1 + \sum_{m=1}^\infty \nu^m e^{-im\theta}  + \sum_{m=1}^\infty \nu^{*m} e^{im\theta}  \right].
	\label{fourier}
\end{equation}
Inserting (\ref{fourier}) and (\ref{vtheta}) into (\ref{cont1}) we obtain the following differential equation for $\nu(t)$:
\begin{equation}
	\dot{\nu} = i \omega \nu -\frac{1}{2} u^* \nu^2 + \frac{1}{2} u
	\label{nudiff}
\end{equation}
where we defined the complex number $u=q e^{i\xi}$. Setting $z=pe^{i \psi}$ and using the definition $\vec{q} = \mathbf{K} \vec{p}$ (see Eq.(\ref{qpar})),
we find
\begin{equation}
	u = K z e^{-i \alpha} - J z^* e^{-i \beta}.
\end{equation}

The last step is to relate the ansatz parameter $\nu$ with the complex order parameter $z$. To do that we note that in the limit of infinitely many
oscillators, Eq.(\ref{paraord}) becomes
\begin{equation}
	z = \int  f(\omega,\theta,t) e^{i\theta} d\theta d\omega.
\end{equation}
Using the Fourier series (\ref{fourier}) we see that only the term proportional to $e^{-i\theta}$ contributes to the integral and we are left with
\begin{equation}
	z = \int  g(\omega) \nu(\omega)  d\omega.
\end{equation}
This equation can be integrated exactly for 
\begin{equation}
	g(\omega) = \frac{\Delta}{\pi} \frac{1}{(\omega-\omega_0)^2 + \Delta^2}
\end{equation}
\cite{Ott2008}. In this case we can write $\nu = \rho e^{i \Phi}$ and we obtain
\begin{equation}
	z = \frac{\Delta}{\pi}  \int  \frac{\rho e^{i\Phi}}{(\omega-\omega_0 + i\Delta)(\omega-\omega_0 - i\Delta)}  d\omega.
\end{equation}
The integral can now be performed in the complex $\omega$-plane using  a closed path from $-R$ to $+R$ over the real line and
closing back from $R$ to $-R$ with a half circle in the positive complex half plane, taking $R \rightarrow \infty$. From Eq.(\ref{nudiff}) we
see that $\Phi$ should be proportional to $\omega$ for large $\omega$, implying that the integral over the half circle goes to zero. Using
the residues theorem at the pole $\omega = \omega_0 + i \Delta$ we obtain
\begin{equation}
	z = \nu(\omega_0 + i \Delta).
\end{equation}

Calculating Eq.(\ref{nudiff}) at $\omega_0 + i \Delta$ allows us to replace $\nu$ by $z$, resulting in
\begin{equation}
	\dot{z} = i (\omega_0 + i \Delta) z -\frac{1}{2} (K z^* e^{i \alpha} - J z e^{i \beta}) z^2 + \frac{1}{2} (K z e^{-i \alpha} - J z^* e^{-i \beta}).
	\label{zdiff}
\end{equation}
Finally, separating real and imaginary parts we obtain equations for the module and phase of the order parameter $z=pe^{i\psi}$:
\begin{equation}
	\dot{p} = -p \Delta + \frac{p}{2}(1-p^2) [K \cos\alpha - J \cos\theta]
	\label{eqp2}
\end{equation}
and
\begin{equation}
	\dot{\theta} = 2 \omega_0 - (1+p^2) [K \sin\alpha - J \sin\theta]
	\label{eqpsi2}
\end{equation}
where we have defined $\theta = 2\psi + \beta$.

\subsection{Phase Diagram}

Non-trivial stationary solutions of Eqs.(\ref{eqp2}) and (\ref{eqpsi2}) are given by
\begin{equation}
	p = \sqrt{1 - \frac{2 \Delta}{K \cos\alpha-J \cos\theta}}
	\label{pequi}
\end{equation}
and
\begin{equation}
	\sin\theta = - \frac{a}{b}
	\label{psiequi}
\end{equation}
where $a=2\omega_0-(1+p^2) K \sin\alpha$ and $b = J(1+p^2)$. These solutions are real provided 
\begin{equation}
	2 \Delta <  K \cos\alpha-J \cos\theta
	\label{condp}
\end{equation}
and $|a| < |b|$, or 
\begin{equation}
	\frac{1}{2}(1+p^2) (K \sin\alpha -|J|) < \omega_0 < \frac{1}{2}(1+p^2) (K \sin\alpha +|J|)
	\label{condpsi}
\end{equation}
Therefore, to find $p$ and $\psi$ for each pair $(\omega_0,\Delta)$ we need to solve Eqs.(\ref{pequi}) and (\ref{psiequi})
and check that conditions (\ref{condp}) and (\ref{condpsi}) are satisfied. 

The trivial (asynchronous) state $p=0$ is always a solution of Eq.(\ref{eqp2}). Although the phase of $z$ for $p=0$ is mostly
irrelevant, it plays a role in analysis of its  stability. At $p=0$ the linearized version of Eq.(\ref{eqp2}) is
\begin{equation}
	\delta \dot{p} = \left(-\Delta + \frac{K \cos\alpha}{2} - \frac{J\cos\theta}{2} \right) \delta p
\end{equation}
whose solution is
\begin{equation}
	\delta p (t) = \exp{ \left\{ -\left(\Delta - \frac{K \cos\alpha}{2}\right)t - \frac{J}{2} \int_0^t \cos\theta(t') dt' \right\} }.
\end{equation}
Therefore, if $\theta$ oscillates (for $\omega_0$ outside the interval in Eq.(\ref{condpsi})) $p=0$ is stable for
$\Delta >  K \cos\alpha/2 \equiv \Delta_0$. When $\theta$ converges to a constant, stability requires
$\Delta >  K \cos\alpha/2 - J \cos\theta/2$. Since the linearized equation for $\theta$ at $p=0$ is $\delta \dot \theta = J \cos\theta \delta \theta$,
for $J>0$, the trivial solution is stable if $\cos\theta < 0$ ($\pi2/ < \theta < 3\pi/2$) and for $J<0$ 
 if $\cos\theta > 0$ ($-\pi/2 < \theta < \pi/2$).

For $\Delta > \Delta_0$ the line separating the async from the static sync region is given in  parametric form 
by $(\omega(\theta), \Delta(\theta))$ with  $\omega(\theta)=(K\sin\alpha -|J| \sin\theta)/2$,  $\Delta(\theta)=(K\cos\alpha - |J|\cos\theta)/2$, 
for $\theta \in (\pi/2,3\pi/2)$.

For $\Delta < \Delta_0$ the solution $p=0$ is unstable and $p$ either converges to a constant (static solutions) or it oscillates (active solutions).  
The boundary between the two kinds of behavior is obtained setting $\sin\theta = \pm1$ and $\cos\theta=0$. From Eq.(\ref{pequi}) we find
\begin{equation}
	\frac{1+p^2}{2} = 1-\frac{\Delta}{K\cos\alpha} = 1 - \frac{\Delta}{2\Delta_0}.
\end{equation} 
Setting $a=b$ we obtain
\begin{equation}
	\frac{1+p^2}{2} = \frac{\omega_0}{K\sin\alpha+|J|} = \frac{\omega_0}{2\omega_{max}}
\end{equation} 
where $\omega_{max} \equiv (K\sin\alpha+|J|)/2$.  Equating these two relations gives the boundary curve
\begin{equation}
	\omega_0(\Delta) = \omega_{max}(2 - \Delta/\Delta_0).
	\label{omegaup}
\end{equation} 
Setting $a=-b$ gives the other boundary curve
\begin{equation}
	\omega_0(\Delta) = \omega_{min}(2 -\Delta/\Delta_0).
	\label{omegadown}
\end{equation} 
where $\omega_{min} \equiv (K\sin\alpha-|J|)/2$. The value of $p$ along the curve is given br Eq.(\ref{pequi}) with
$\cos\theta=0$, and it approaches 1 as $\Delta \rightarrow 0$. The phase is $\psi = 3\pi/4-\beta$ on the upper curve and $\psi=\pi/4 - \beta$ on the lower curve.
Between these two curves the solution is static and outside this interval the
phase $\theta$, and therefore $\psi$, oscillates, leading to an oscillation of $p(t)$, corresponding to active states.

\begin{figure}
	\includegraphics[scale=0.29]{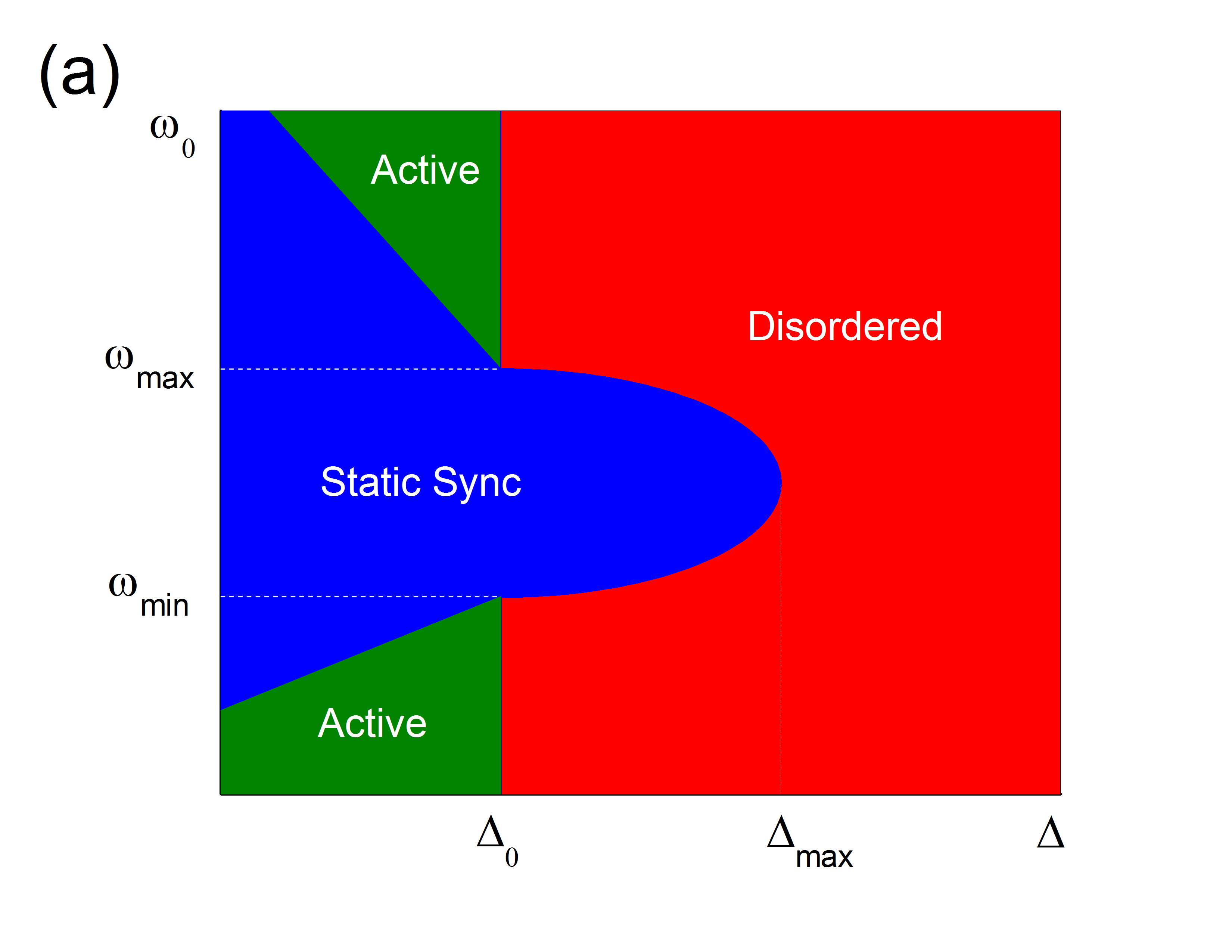} 
	\includegraphics[scale=0.29]{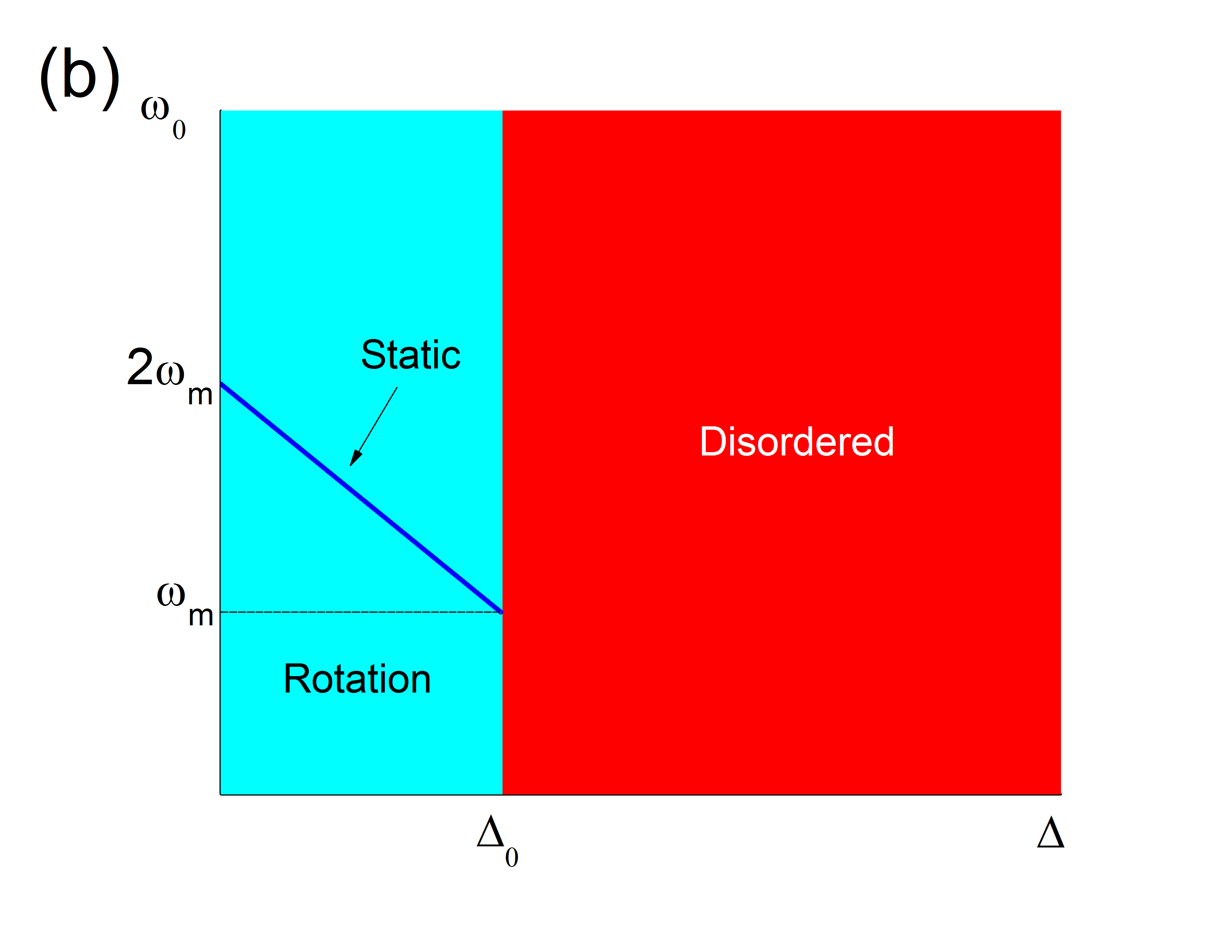} 
	\caption{Phase diagram in the $\omega_0 \times \Delta$ plane for $D=2$. (a) The static sync region is bounded by $\omega_{min}=(K\sin\alpha -|J| )/2$ 
		and $\omega_{max}=(K\sin\alpha + |J| )/2$,with $\Delta_0=K\cos\alpha/2$ and $\Delta_{max} = \Delta_0 + |J|/2$. The curved line separating
		the async from the static sync region is given in  parametric form by $(\omega(\theta), \Delta(\theta))$ with 
		 $\omega(\theta)=(K\sin\alpha -|J| \sin\theta)/2$,  $\Delta(\theta)=(K\cos\alpha - |J|\cos\theta)/2$, for $\theta \in (\pi/2,3\pi/2).$ (b) For $J=0$
		 the Kuramoto-Sakaguchi model is recovered and the order parameter rotates if $K>2\Delta/\cos\alpha$. Rotation changes direction as the
		 blue line is crossed, with $\omega_m = K \sin\alpha/2$.  }
	\label{fig1}
\end{figure}

For $J=0$ the static sync region collapses and the active regions become regions of rotation, where the module of $p$ remains constant
whereas the phase $\psi$ rotates with angular velocity $\omega_0-K\sin\alpha+\Delta \tan\alpha$. A line of static sync appears for 
$\omega_0 = K\sin\alpha  - \Delta \tan\alpha$. The full diagram is illustrated in Figure \ref{fig1}.

\section{Simulations}
\label{sim}

In this section we present simulations of the generalized Kuramoto model for $D=2$, 3 and 4 for the coupling matrices described
in section \ref{rep}. For each coupling matrix and distribution of natural frequencies we integrate the system for different values of 
$\omega_0$ and  $\Delta$.  In order to characterize the asymptotic state of the system we computed the following quantities: 
(i) $\langle p \rangle$, time average of the module of the order parameter ; (ii) $\delta p$, mean square deviation around the average, 
and; (iii) $\delta_{max}$, maximum between mean square deviation of the components of $\vec{p}$. The first quantity informs about the degree of 
synchronization whereas the second indicates if $p$ is constant or displays oscillatory motion (active state). Finally, $\delta_{max}$ indicates 
if $\vec{p}$ is rotating ($\delta_{max} > 0$) or not ($\delta_{max}=0$).  

Capital letters in the phase diagrams indicate the asymptotic state of the system as follows:

D - disordered (not synchronized) - $\langle p \rangle \approx 0$.

S - static sync - $\langle p \rangle > 0$, $\delta p \approx 0$, $\delta_{max} \approx 0$. 

R - rotation - $\langle p \rangle > 0$, $\delta p \approx 0$, $\delta_{max} > 0$. 

A - active sync - $\langle p \rangle > 0$, $\delta p > 0$, $\delta_{max} > 0$.

In all simulations we set $N=5000$ oscillators for a total integration time of $t=2000$, using the last $500$ units of time
to compute the asymptotic results. The only exception is the case of the Lorentz distribution, for which we used $N=200$. $t=1000$
and the last $250$ units of time for averages, due to the very slow convergence of the system. Integration of the equations of motion
were performed with a 4th order Runge-Kutta algorithm with precision parameter $10^{-6}$. Convergence of the results was also 
checked against the method proposed in \cite{de2023numerical}. Initial conditions for the oscillators were chosen randomly on the sphere. 
The parameters $\omega_0$ and $\Delta$ in heat maps were varied from 0 to 2 at steps of 0.05.

\subsection{Phase diagrams in $D=2$}

Although in $D=2$ we have the exact phase diagrams for the Lorentz distributions, simulations are important for two reasons:
first, we don't have the diagrams for other distributions; second, since we need to automatize the simulations for other distributions
and higher dimensions, we need to make sure we can extract the different phases from the simulated diagrams. Therefore, we first
simulate the diagrams for the Lorentz distribution
\begin{equation}
	g_L(\omega) = \frac{\Delta}{\pi} \frac{1}{(\omega-\omega_0)^2 + \Delta^2};
	\label{glor}
\end{equation}
and see how they compare with the exact solutions. We also simulated the equations for the Gaussian
\begin{equation}
	g_G(\omega) = \frac{1}{\sqrt{2\pi\Delta^2}} e^{-\frac{(\omega-\omega_0)^2}{2\Delta^2}};
	\label{ggauss}
\end{equation}
and uniform distributions,
\begin{equation}
	g_U(\omega) = \left\{
	\begin{array}{l}
		\frac{1}{\Delta} \qquad {\mbox if}  -\frac{\Delta}{2} + \omega_0 \leq \omega \leq \omega_0 + \frac{\Delta}{2}  \\
		0 \qquad {\mbox{otherwise}}.
	\end{array}
	\right.
	\label{guni}
\end{equation}
Note that $\Delta$ is a measure of the width of the distributions, but has different meanings in each case. For the Gaussian
distribution $\Delta^2$ is the variance; for the uniform distribution the variance is $\Delta^2/12$ whereas the Lorentz has infinite
variance.

For each distribution we simulated the dynamics with three coupling matrices (see Eq.(\ref{k2D})):
\begin{equation}
	\mathbf{K} _S = \left(
	\begin{array}{cc}
		2.5 \quad &  0 \\
		0 \quad & 0.8 
	\end{array}
	\right)
	\label{2d1}
\end{equation}
\begin{equation}
	\mathbf{K} _R = 2. 5 \left(
	\begin{array}{cc}
		\cos0.5 \quad &  \sin0.5 \\
		-\sin0.5 \quad & \cos0.5 
	\end{array}
	\right)
	\label{2d2}
\end{equation}
and
\begin{equation}
	\mathbf{K} _A = \mathbf{K} _R + \left(
	\begin{array}{cc}
		0.2 \quad &  0.1 \\
		0 \quad & 0 
	\end{array}
	\right).
	\label{2d3}
\end{equation}
These choices correspond to coupling matrices with real eigenvalues, leading to static states for $\Delta=\omega_0=0$ ($\mathbf{K} _S$), purely complex eigenvalues, leading to rotations as in the Kuramoto-Sakaguchi model  ($\mathbf{K} _R$) and complex eigenvalues corresponding
to active states ($\mathbf{K} _A$).

\begin{figure}
	\includegraphics[scale=0.4]{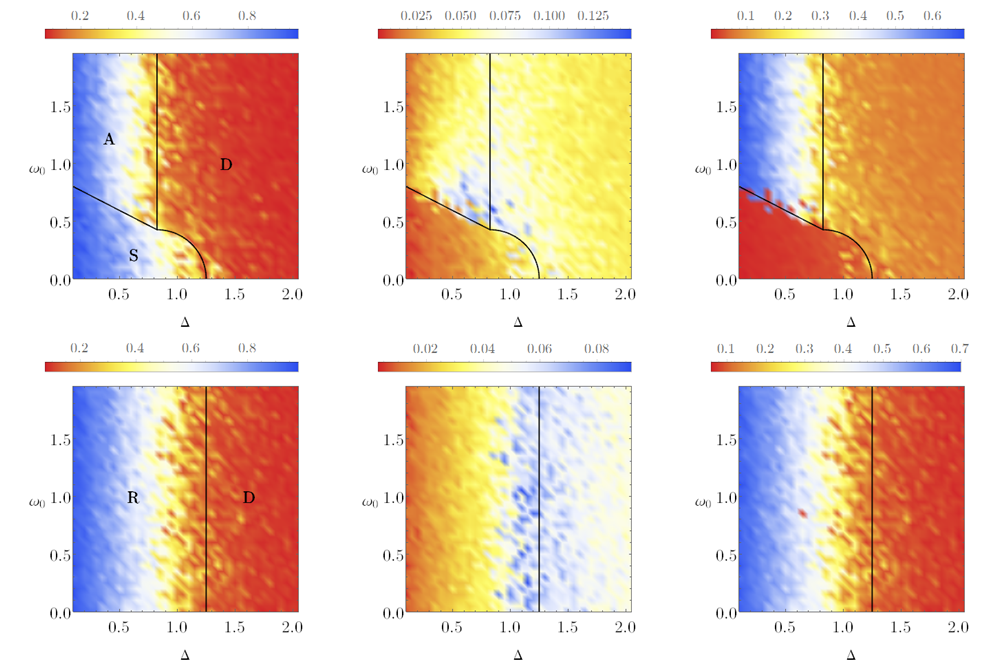}
	\caption{Heat maps in the $\omega_0$-$\Delta$ plane for the Lorentz distribution. Panels show results for 
		coupling matrices $\mathbf{K}_S$ (first line) and $\mathbf{K}_R$ (second line). Along each line of plots, panels
		show the average value of the order parameter $\langle p \rangle$, its mean square deviation $\delta p$ and 
		$\delta_{max}$. Black lines show the theoretical results (see Fig.\ref{fig1}).}
	\label{fig2}
\end{figure}

\begin{figure}
	\includegraphics[scale=0.4]{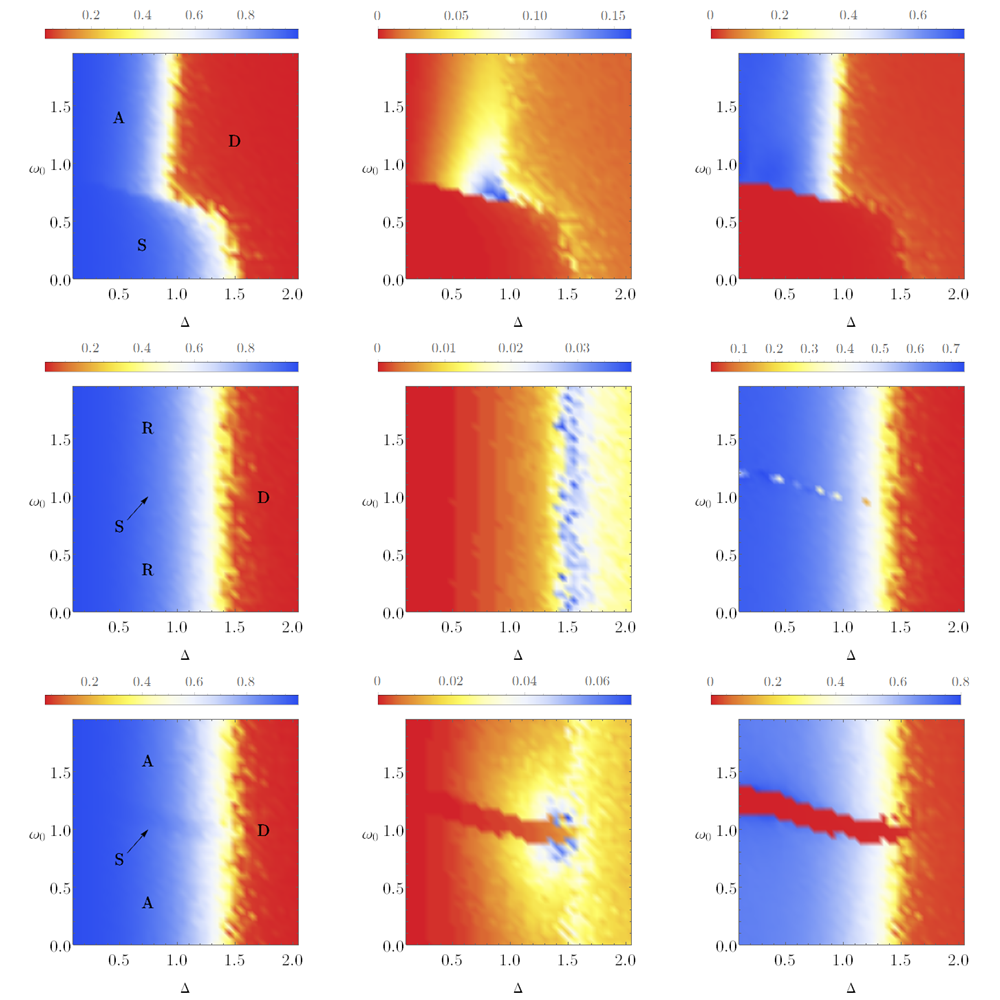}
	\caption{Heat maps in the $\omega_0$-$\Delta$ plane for the Gaussian distribution. Panels show results for 
		coupling matrices $\mathbf{K}_S$ (first line) , $\mathbf{K}_R$ (second line) and $\mathbf{K}_A$ (third line). 
		Each line shows the average value of the order parameter $\langle p \rangle$, its mean square deviation $\delta p$ and 
		$\delta_{max}$.}
	\label{fig3}
\end{figure}

\begin{figure}
	\includegraphics[scale=0.4]{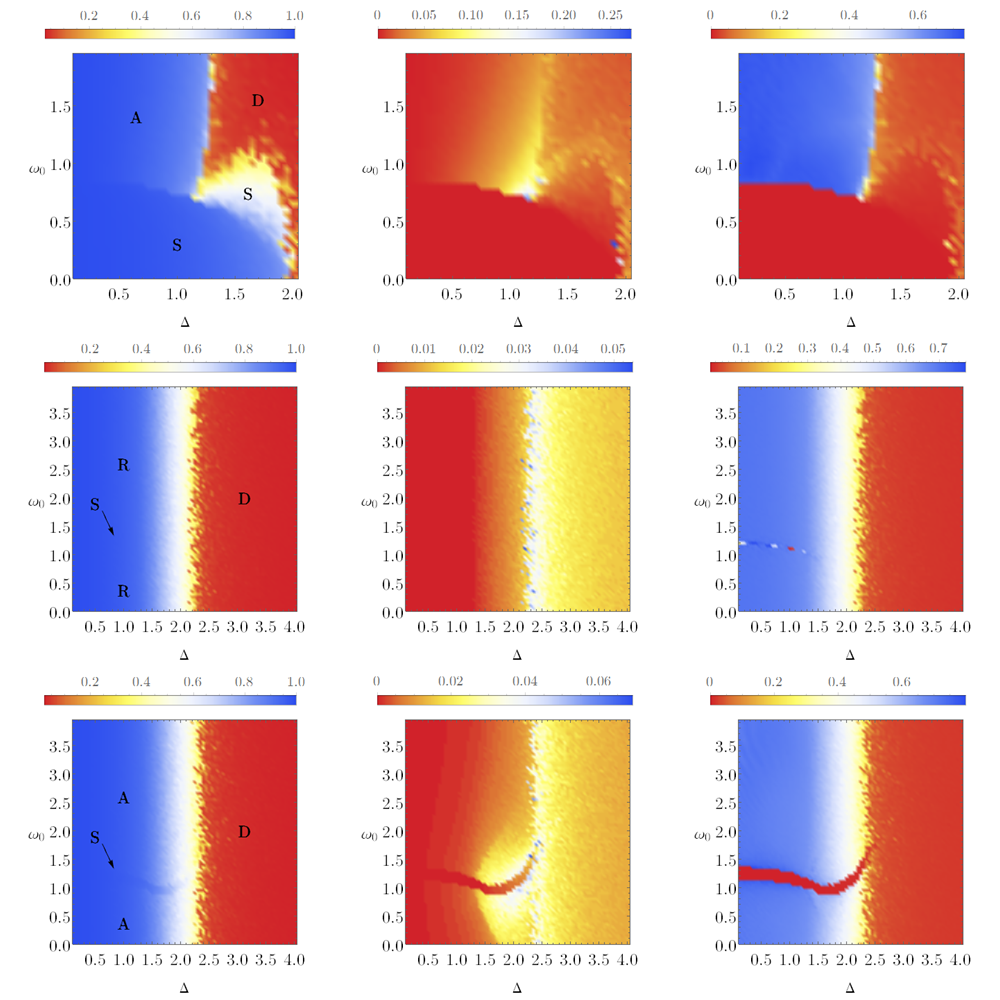}
	\caption{Heat maps in the $\omega_0$-$\Delta$ plane for the uniform distribution.  Panels show results for 
		coupling matrices $\mathbf{K}_S$ (first line) , $\mathbf{K}_R$ (second line) and $\mathbf{K}_A$ (third line). 
		Plots show the average value of the order parameter $\langle p \rangle$, its mean square deviation $\delta p$ and 
		$\delta_{max}$.}
	\label{fig4}
\end{figure}

Fig.~\ref{fig2} displays results for the Lorentz distribution, which we simulate only for $\mathbf{K} _S$ (top panels) and $\mathbf{K} _R$ 
(lower panels), as our intention is to validate the numerical results comparing with the exact diagrams. Continuous black lines show the boundary 
curves as in Fig.\ref{fig1} and they divide the plane int o three (top) or two (bottom) regions. In both cases the right most region corresponds
to non-synchronized states. The top panels show static synchronization, with $\vec{p}$ constant, in the lower left corner, as can be seen by the low 
values of both $\delta p$ and $\delta_{max}$. The upper left region, on the other hand, shows active states, with $\langle p \rangle$ constant but 
rotating and oscillating $\vec{p}(t)$, as indicated by  significant values of $\delta p$ and large values of $\delta_{max}$.
For the bottom panels, corresponding to the Kuramoto-Sakaguchi model, $\vec{p}$ rotates but keeps its module constant ($\delta p \approx 0$).

Fig.~\ref{fig3} shows similar plots for the Gaussian distribution, Eq.(\ref{ggauss}). The top two panels are qualitatively similar to the panels in 
Fig.\ref{fig2}, although not identical: critical transition values of $\Delta$ and $\omega_0$ are different and transitions are much sharper, as 
natural frequencies far from $\omega_0$ are much less likely to be sampled. Interestingly, for the case of $\mathbf{K}_A$, the line of fixed 
$\vec{p}$ immersed in the rotating zone (see Fig.\ref{fig1}(b) ) is enlarged into an area (red stripe in the $\delta_{mas})$, showing the great 
sensitivity of phase diagram with the coupling matrix.

Finally, Fig.~\ref{fig4} shows results for the uniform distribution, Eq.(\ref{guni}) and the same coupling matrices, Eqs.(\ref{2d1}) -(\ref{2d3}). 
Except for the case of  $\mathbf{K}_R$  (middle row) which is qualitatively similar to the cases of Lorentz and Gaussian distributions, 
new regions develop in this case. For $\mathbf{K}_S$ part of the region corresponding to non-synchronized states for Gaussian and 
Lorentz distributions becomes partially synchronized  with static states (first row) and for $\mathbf{K}_A$ the active states near the line
of static states develop larger oscillations (yellow area in the middle plot in the third row).  Enlargement of the line of static sync states
is also observed in this case.

% %%%%%%%%%%%%%%%%%%%%%%%%%%%%%%%%%%%%%%%%%%%%%%%%%%%%
% %%%%%%%%%%%%%%%%%%%%%%%%%%%%%%%%%%%%%%%%%%%%%%%%%%%%
\subsection{Phase diagrams in $D=3$}

%d1: Uniform angular and Gaussian module centered at omegamod
%d2: Entries are Gaussian distributed around omegamod
%d3: Entries are uniformly distributed around omegamod

% I- c1 -- a=1, b=0.5,  alpha=0.5
% II- c2 -- a=1, b=1,  alpha=0.5

% 3d = matrix K - a for rotations with alpha, b for real eigenvalue
% wmat3 = matrix Wi
% vecwmat3 = omega_i

To explore the phase diagrams in 3D we set the coupling matrix as in Eq.(\ref{3d}) with $a=1$ and $\alpha=0.5$ and consider
two values of the parameter $b$. For $b=0.5$ we define
\begin{equation}
	\mathbf{K}_{3R} = \left(
	\begin{array}{ccc}
		\cos 0.5 & \sin 0.5 & 0 \\
		-\sin 0.5 & \cos 0.5 & 0 \\
		0 & 0 & 0.5
	\end{array}
	\right) 
	\label{3da}
\end{equation}
and for $b=1$
\begin{equation}
	\mathbf{K}_{3S} = \left(
	\begin{array}{ccc}
		\cos 0.5 & \sin 0.5 & 0 \\
		-\sin 0.5 & \cos 0.5 & 0 \\
		0 & 0 & 1
	\end{array}
	\right).
	\label{3da}
\end{equation}
In the first case the dominant eigenvalues have complex eigenvectors in the $\hat{e}_1$-$\hat{e}_2$ plane, corresponding to rotations
for $\Delta=\omega_0=0$. In the second case the dominant eigenvector is real in the $\hat{e}_3$ direction leading to static synchronization. 
For each case we use three different types of natural frequencies  distributions, described either in terms of the matrix $\mathbf{W}_i$ entries 
as in Eq.(\ref{3d}) or in terms of the associated vector $\vec{\omega}_i$, Eq.(\ref{vecwmat3}), as follows:

\noindent -- $g_{ang}(\vec{\omega})$; for each oscillator a vector $\vec{\omega}_i$ is sampled with uniform distribution of angles $\alpha_i$
and $\beta_i$ and Gaussian distribution of module $\omega_i$ centered at $\omega_0$ with width $\Delta$.

\noindent -- $g_{gauss}(\vec{\omega})$; each entry of $\mathbf{W}_i$ is sampled from a Gaussian distribution centered at $\omega_0$ 
with width $\Delta$.

\noindent -- $g_{uni}(\vec{\omega})$; each entry of $\mathbf{W}_i$ is sampled from a uniform distribution centered at $\omega_0$ with
width $\Delta$.\\

\begin{figure}
	\includegraphics[scale=0.4]{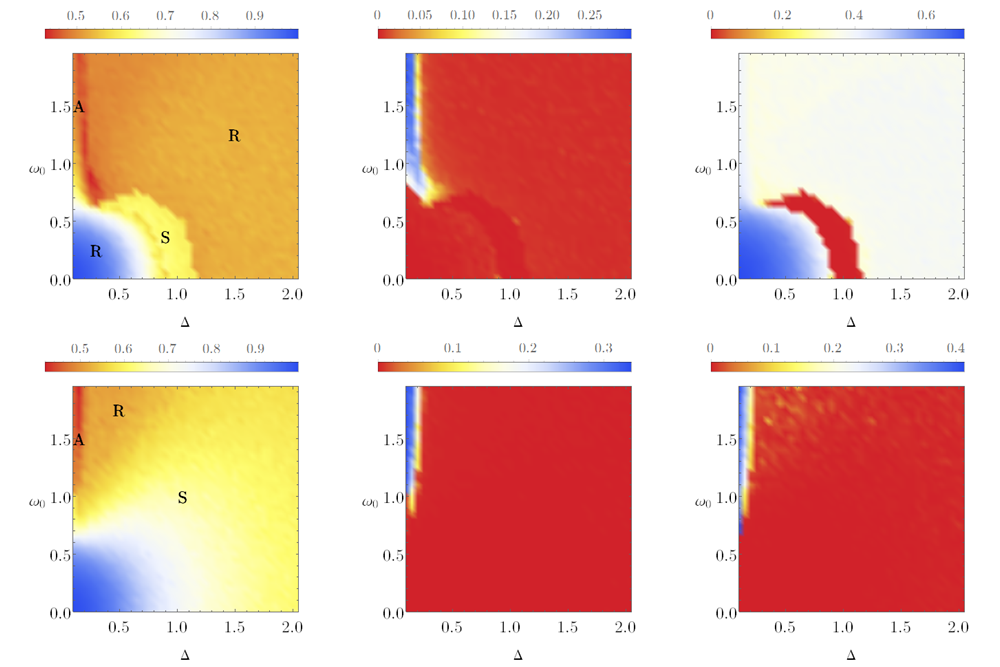}
	\caption{Heat maps in the $\omega_0$-$\Delta$ plane for 3D and distribution $g_{ang}(\vec{\omega})$. Panels show results for 
		coupling matrices $\mathbf{K}$ as in Eq.(\ref{3d}) with $a=1$, $\alpha=0.5$. The value of $b$ is 0.5 (first line) and 1.0 (second line). 
		Plots show the average value of the order parameter, its mean square deviation and $\delta_{max}$. }
	\label{fig5}
\end{figure}

\begin{figure}
	\includegraphics[scale=0.4]{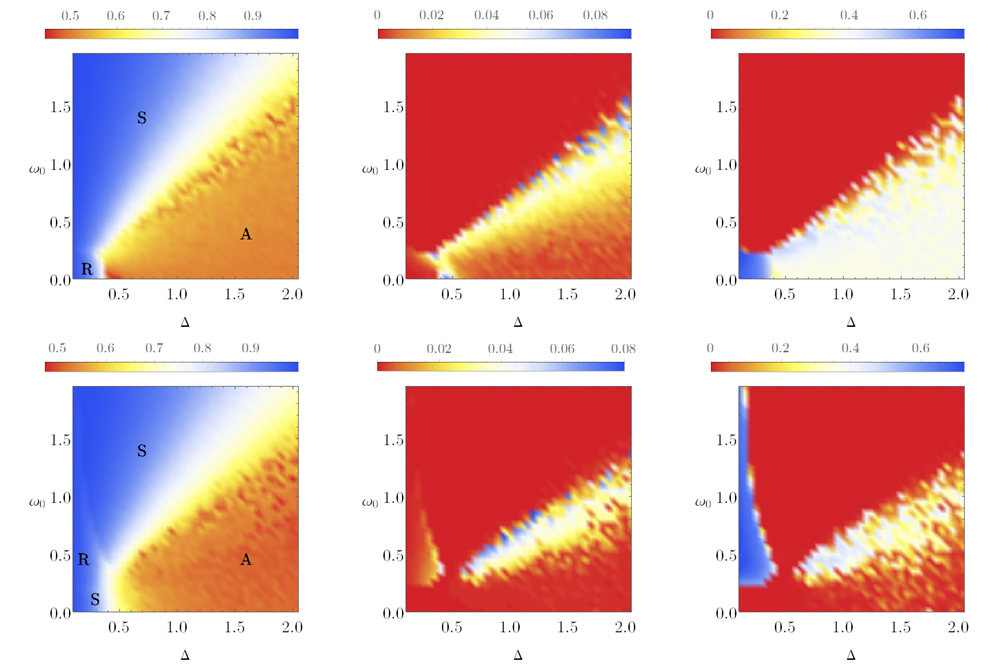}
	\caption{Heat maps in the $\omega_0$-$\Delta$ plane for the 3D case. Here $g(\vec{\omega} )$ is given by  Eq,(\ref{wmat3}) with all
		entries Gaussian distributed around $\omega=1$.  Panels show results for 
		coupling matrices $\mathbf{K}$ as in Eq.(\ref{3d}) with $a=1$, $\alpha=0.5$. The value of $b$ is 0.5 (first line) and 1.0 (second line). 
		Plots show the average value of the order parameter, its mean square deviation and $\delta_{max}$. }
	\label{fig6}
\end{figure}

\begin{figure}
	\includegraphics[scale=0.4]{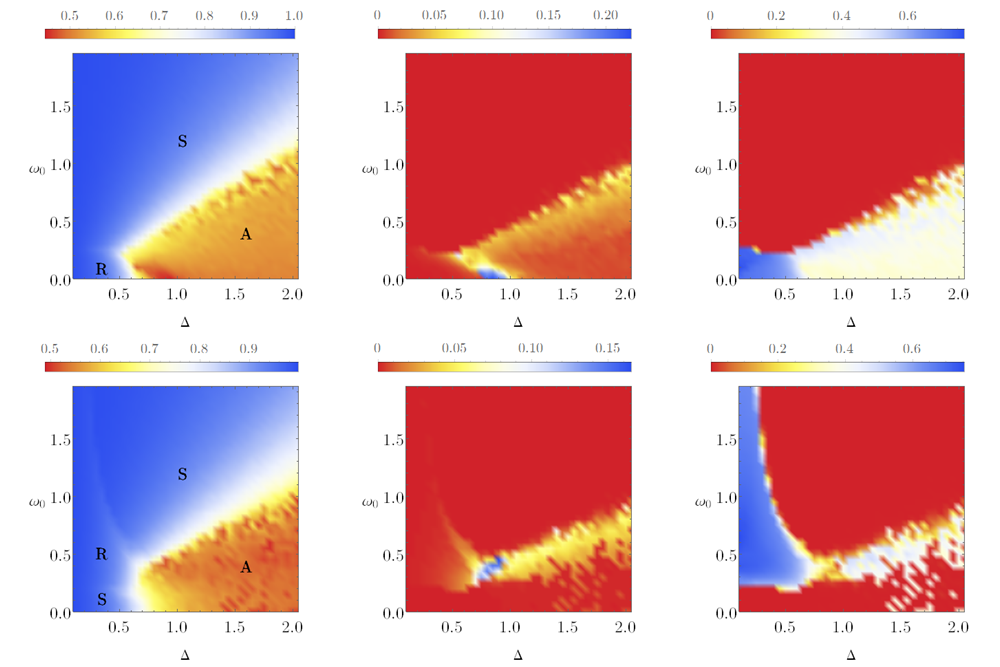}
	\caption{Heat maps in the $\omega_0$-$\Delta$ plane for the 3D case. Here $g(\vec{\omega} )$ is given by  Eq,(\ref{wmat3}) with all
		entries uniformly distributed around $\omega=1$.  Panels show results for 
		coupling matrices $\mathbf{K}$ as in Eq.(\ref{3d}) with $a=1$, $\alpha=0.5$. The value of $b$ is 0.5 (first line) and 1.0 (second line). 
		Plots show the average value of the order parameter, its mean square deviation and $\delta_{max}$. }
	\label{fig7}
\end{figure}

Figs.~\ref{fig5} to \ref{fig7} show results of numerical simulations in each case. As in the 2D case we show the time-average value of the 
order parameter after the transient, $\langle p \rangle$, its mean square deviation $\delta p$, and $\delta_{max}$, the maximum between 
mean square deviation of the components of $\vec{p}$, indicating if $\vec{p}$ is rotating ($\delta_{max} > 0$) or not ($\delta_{max}=0$). 
Capital letters indicate the asymptotic state of the system as in $D=2$. In all cases, when $\Delta$ and $\omega_0$ are sufficiently small the 
oscillators synchronize and rotate (case 1, first line in all figures) or converge to a static configuration (case 2, second line of figures). However, 
as $\Delta$ and $\omega_0$ increase, the behavior of the system depends significantly on the distribution of natural frequencies. 

For $g_{ang}(\vec{\omega})$, synchronization decreases as $\Delta$ and $\omega_0$ increase. For the coupling matrix $\mathbf{K}_{3R}$
a phase transition from rotation (R) to static sync (S) and back to rotation (R) is observed  as $\Delta$ and $\omega_0$ increase. A thin region
of active states is also noted for small $\Delta$ and large $\omega_0$. For $\mathbf{K}_{3S}$ the diagram is dominated by a large area 
of static sync (S), although a similar region of active states is observed, next to rotations (R). Notice that, since the direction of the vectors
$\vec\omega_i$ is uniformly sampled, the average value of these vectors is zero for $g_{ang}$, independent of the value of $\omega_0$. 
This makes the phases observed at $\omega_0=\Delta=0$ to extend over large regions of the diagram as compared to the other two distributions
in Figs.~\ref{fig6} and \ref{fig7}. 

The phase diagrams for $g_{gauss}$ and $g_{uni}$ are similar, but very different from that of $g_{ang}$. Synchronization is much facilitated
in these cases, as we don't see regions of disordered motion in this range of parameters. Moreover, states with nearly complete sync are possible
even for large $\Delta$ in the static phase. The phase of active states (A) is also much larger than in Fig.~\ref{fig5} and occurs for small values of 
$\omega_0$ and large values of $\Delta$. Pure rotations tend to be suppressed for $g_{gauss}$, occurring in small regions for both
$\mathbf{K}_{3R}$ and $\mathbf{K}_{3S}$ and in a larger region $g_{uni}$, especially for $\mathbf{K}_{3S}$. Finally we note that the 
uniform distribution, Fig.~\ref{fig7}, produces sharper transitions between  the different phases.

% %%%%%%%%%%%%%%%%%%%%%%%%%%%%%%%%%%%%%%%%%%%%%%%%%%%%
% %%%%%%%%%%%%%%%%%%%%%%%%%%%%%%%%%%%%%%%%%%%%%%%%%%%%
\subsection{Phase diagrams in $D=4$}

%d1: Uniform angular for (w1,w2,w3) and (w4,w5,w6) and Gaussian module centered at omegamod   - fig8
%d2: Entries are Gaussian distributed around omegamod                    - fig9
%d3: Entries are uniformly distributed around omegamod                     - fig10

%I- c1: alpha=0, a1=2.5, a2=0.8, b=0.5, beta=0.5
%II- c2: alpha=0, a1=0.5, a2=0.8, b=2.5, beta=0.5

%K11 = a1*cos(alpha)    K12 = a1*sin(alpha)    K13 = 0                        K14 = 0
%K21 = -a2*sin(alpha)    K22 = a2*cos(alpha)   K23 = 0                        K24 = 0
%K31 = 0                         K32 = 0                       K33 = b*cos(beta)       K34 = b*sin(beta)
%K41 = 0                         K42 = 0                        K43 = -b*sin(beta)      K44 = b*cos(beta)

In this section we illustrate the phase diagrams in $D=4$ with  two instances of the coupling matrix Eq.~(\ref{4d}). In both cases
we fixed $\alpha=0$, $\beta=0.5$ and $a_2=0.8$. Similar to the 3D system we choose the remaining parameters $a_1$ and $b$ 
as follows: first we set $a_1=2.5$ and $b=0.5$, so that the leading eigenvalue of $\mathbf{K}$ is real in the direction of $\hat{e}_1$:
\begin{equation}
	\mathbf{K}_{4S} = \left(
	\begin{array}{cccc}
		2.5  & 0 & 0 & 0\\
		0 & 0.8  & 0 & 0\\
		0 & 0 & 0.5 \cos 0.5 & 0.5 \sin 0.5 \\
		0 & 0 & -0.5 \sin 0.5 & 0.5 \cos 0.5 
	\end{array}
	\right). 
	\label{4ds}
\end{equation}

For the second case we set $a_1=0.5$ and $b=2.5$, with a complex leading eigenvalue in the $\hat{e}_3$-$\hat{e}_4$ plane:
\begin{equation}
	\mathbf{K}_{4R} = \left(
	\begin{array}{cccc}
		0.5  & 0 & 0 & 0\\
		0 & 0.8  & 0 & 0\\
		0 & 0 & 2.5 \cos 0.5 & 2.5 \sin 0.5 \\
		0 & 0 & -2.5 \sin 0.5 & 2.5 \cos 0.5 
	\end{array}
	\right) 
	\label{4dr}
\end{equation}

\begin{figure}
	\includegraphics[scale=0.4]{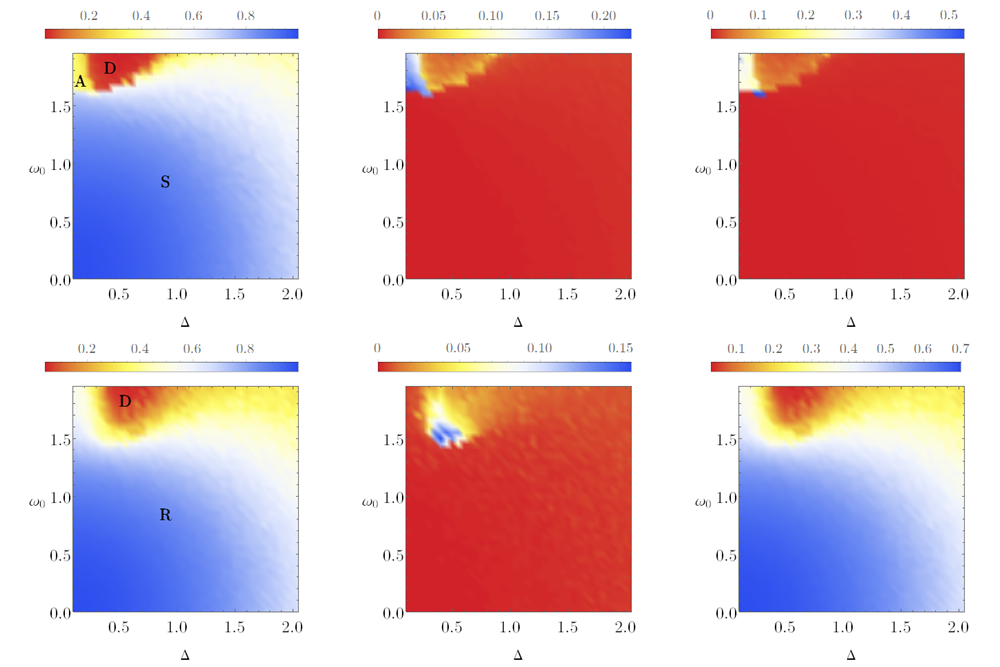}
	\caption{Heat maps in the $\omega_0$-$\Delta$ plane for the 4D case with distribution of natural frequencies $g_{ang4}$. 
		Panels show results for coupling matrices $\mathbf{K}_{4S}$ (first line) and $\mathbf{K}_{4R}$ (second line), as in 
		Eqs.(\ref{4ds}) and (\ref{4dr}). Plots show the average value of the order parameter, its mean square deviation and $\delta_{max}$. }
	\label{fig8}
\end{figure}

\begin{figure}
	\includegraphics[scale=0.4]{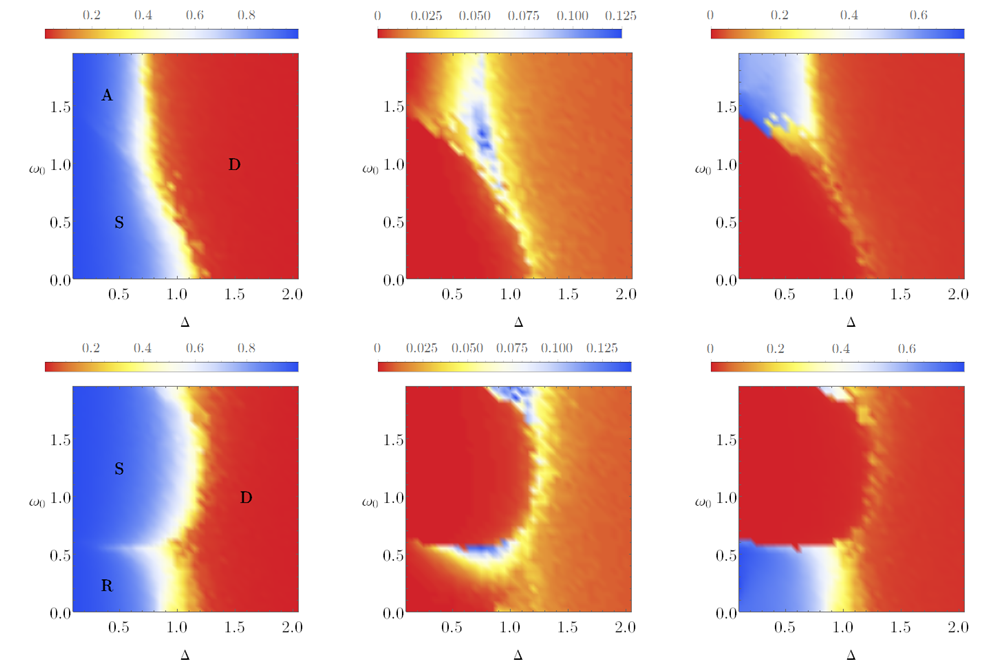}
	\caption{Heat maps in the $\omega_0$-$\Delta$ plane for the 4D case with distribution of natural frequencies $g_{gauss4}$. 
		Panels show results for coupling matrices $\mathbf{K}_{4S}$ (first line) and $\mathbf{K}_{4R}$ (second line), as in 
		Eqs.(\ref{4ds}) and (\ref{4dr}). Plots show the average value of the order parameter, its mean square deviation and $\delta_{max}$.  }
	\label{fig9}
\end{figure}

\begin{figure}
	\includegraphics[scale=0.4]{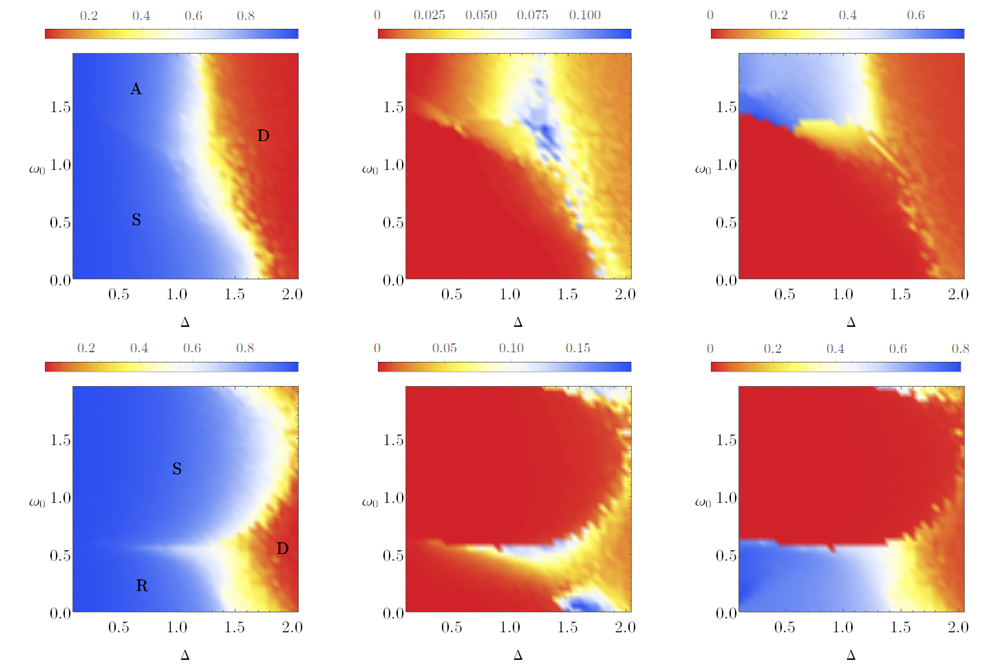}
	\caption{Heat maps in the $\omega_0$-$\Delta$ plane for the 4D case with distribution of natural frequencies $g_{uni4}$. 
		Panels show results for coupling matrices $\mathbf{K}_{4S}$ (first line) and $\mathbf{K}_{4R}$ (second line), as in 
		Eqs.(\ref{4ds}) and (\ref{4dr}). Plots show the average value of the order parameter, its mean square deviation and $\delta_{max}$. }
	\label{fig10}
\end{figure}

In $D=4$ there are six independent entries for each matrix of natural frequencies $\mathbf{W}_i$ and many ways to
choose these values.  For each choice of coupling matrix above we performed simulations for the following three distributions:

\noindent -- $g_{ang4}$; in order to have a distribution similar to that used in Fig.~\ref{fig5}, we grouped the six entries $\omega_{ji}$
for each oscillator $i$ into two vectors $\vec{\xi}_1 = (\omega_{1i},\omega_{2i},\omega_{3i})$ and $\vec{\xi}_2 = (\omega_{4i},\omega_{5i},\omega_{6i})$
and sampled $\vec{\xi}_1$ and $\vec{\xi}_2$ with uniform angular distribution and Gaussian distribution of modules $\xi_1$ and $\xi_2$, 
centered at $\omega_0$.

\noindent -- $g_{gauss4}$; all entries are Gaussian distributed around $\omega_0$.

\noindent -- $g_{uni4}$; all entries are uniformly distributed around $\omega_0$.\\

Fig.~\ref{fig8} shows results for $g_{ang4}$. Similar to the 3D case with $g_{ang}$, the average of the natural frequencies is zero 
for all $\omega_0$ and $\Delta$, since the entries of $\mathbf{W}_i$ are uniformly distributed in all directions, with only its average intensity
controlled by $\omega_0$. The basic phases S and R dominate much of the phase diagrams for $\mathbf{K}_{4S}$ and $\mathbf{K}_{4R}$
respectively. 

For $g_{gauss4}$ and $g_{uni4}$ we obtain qualitatively similar diagrams (Figs.~\ref{fig9} and \ref{fig10}), where rotations are 
suppressed for $\mathbf{K}_{4S}$ and 
active states are suppressed for  $\mathbf{K}_{4R}$. The phase diagrams are very different from their 3D counterparts exhibited in
Figs.~\ref{fig6} and \ref{fig7}, but somewhat similar to the 2D case, Figs.~\ref{fig3} and \ref{fig4}, highlighting the role of dimensional
parity (even or odd) in the dynamics and equilibrium properties of the model  \cite{chandra2019continuous}.
Synchronization happens only for limited values of $\Delta$, especially for the Gaussian case.

% %%%%%%%%%%%%%%%%%%%%%%%%%%%%%%%%%%%%%%%%%%%%%%%%%%%%
% %%%%%%%%%%%%%%%%%%%%%%%%%%%%%%%%%%%%%%%%%%%%%%%%%%%%
\section{Conclusions}

The multidimensional Kuramoto model was proposed in \cite{chandra2019continuous} as a natural extension of the
original system of coupled oscillators. In the extended model, oscillators are first reinterpreted as 
interacting particles moving on the unit circle. The system is then generalized allowing the particles to move on the surface of unit
spheres embedded in D-dimensional spaces. For $D=2$ the original model is recovered. The equations of motion of the multidimensional
model,  Eq.~(\ref{eq3}), are formally identical in any number of dimensions, provided they are written in terms of the unit vectors determining 
the particles' positions in the corresponding space.

The vector form of equations (\ref{eq3}) describing the multidimensional model admits a further natural extension, where the coupling
constant is replaced by a coupling matrix as in Eq.~(\ref{kuragen}), breaking the rotational symmetry and promoting generalized frustration
between the particles \cite{buzanello2022matrix,de2023generalized}. For identical oscillators, when all natural frequencies are set to zero,
the asymptotic dynamic is completely determined by the eigenvectors and eigenvalues of the coupling matrix $\mathbf{K}$. If the leading
eigenvalue is real and positive the order parameter $\vec{p}$ converges to $p=1$ in the direction of the eigenvector (static sync). If the leading 
eigenvector is complex, $\vec{p}$ rotates in the plane defined by the real and imaginary parts of the corresponding eigenvector, also with $p=1$. 
These results hold for all dimensions $D$. 

Here we have shown that this simple behavior changes dramatically for non-identical oscillators. The asymptotic nature of the system depends
strongly on the distribution of natural frequencies, on the coupling matrix and on the dimension $D$.  In order to simplify the calculations we 
parametrized the distributions of natural frequencies by only two quantities related to their average value, 
$\omega_0$, and width $\Delta$.  Because the coupling matrix breaks the rotational symmetry, the magnitude of $\omega_0$ plays a
key role in the dynamics. We constructed phase diagrams in the $\omega_0 \times \Delta$ plane for different types of distributions and for 
dimensions 2, 3 and 4. In the case of $D=2$ we computed the phase diagram analytically for the Lorentz distribution and numerically for the Gaussian 
and uniform distributions. 

In $D=3$ synchronization starts at $p=0.5$ if the real part of the leading eigenvalue of $\mathbf{K}$ is positive \cite{chandra2019continuous,barioni2021complexity} and the phase diagram exhibits all phases: static sync, rotation and active states. The size
and disposition of each phase changes according to the coupling matrix and distribution $g(\vec{\omega})$. All phase diagrams in $D=3$ are
remarkably different from their counterparts $D=2$.  Finally, for $D=4$ the phase diagrams have structures similar to their equivalents in $D=2$, 
showing that the parity of $D$ matters as in the case of diagonal coupling matrices \cite{chandra2019continuous}. 

As active states prevent full synchronization of the particles, knowledge of their location in parameter space is an important information.
In general terms we can say that 3D systems are characterized by large regions of active states that appear for small values of 
$\omega_0$ and large values of $\Delta$. For even dimensional systems, on the other hand, active states require large values of
$\omega_0$ and small values of $\Delta$.

\begin{acknowledgments}
	This work was partly supported by FAPESP, grant 2021/14335-0 (ICTP‐SAIFR) and CNPq, grant 301082/2019‐7. 
\end{acknowledgments}

%%%%%%%%%%%%%%%%%%%%%%%%%%%%%%%%%%%%%%%%%%%%%%%%%%%%%
\clearpage 
\newpage
\bibliographystyle{ieeetr}
%\bibliography{kuramoto.bib}

\end{document}